\documentclass[reprint]{revtex4-2}

\usepackage{graphicx}% Include figure files
\usepackage{dcolumn}% Align table columns on decimal point
\usepackage{bm}% bold math
\usepackage[hidelinks, linktocpage=true]{hyperref}
\usepackage{amsmath,amssymb,amsfonts}
\usepackage{algorithmic}
\usepackage{graphicx}
\usepackage{textcomp}
\usepackage{xcolor}
\usepackage{color}
\usepackage{centernot}
\usepackage{soul}

\definecolor{forestgreen}{rgb}{0.13, 0.55, 0.13}

\newcommand{\blankout}[1]{}
\newcommand{\Xd}{X_d}

\newcommand{\Pspace}{{\cal P}}
\newcommand{\Fspace}{{\cal L}}

\newcommand{\Dchannel}{\Delta\!-\!\mbox{channel}}

\newcommand{\be}{\begin{equation}}
\newcommand{\ee}{\end{equation}}

\DeclareMathOperator*{\argmax}{argmax}

\begin{document}

\title{Evaluating evolution as a learning algorithm}

\author{Miles Miller-Dickson}
 \email{miles\_miller-dickson@brown.edu}
\author{Christopher Rose}%
 \email{christopher\_rose@brown.edu}
\affiliation{%
Department of Electrical Engineering, \\
Brown University, Providence, RI, 02906
}%

\author{C. Brandon Ogbunugafor}
 \email{brandon.ogbunu@yale.edu}
\affiliation{
Department of Ecology and Evolutionary Biology,\\ 
Yale University, New Haven, CT, 06520 
}%
\author{I. Saira Mian}
\email{s.mian@cs.ucl.ac.uk}
\affiliation{%
 Department of Computer Science,\\
 University College London,\\ London, U.K.
}%

\date{\today}

\begin{abstract}
We interpret the Moran model of natural selection and drift as an algorithm for learning features of a simplified fitness landscape, specifically genotype superiority. This algorithm's efficiency in extracting these characteristics is evaluated by comparing it to a novel Bayesian learning algorithm developed using information-theoretic tools. This algorithm makes use of a communication channel analogy between an environment and an evolving population. We use the associated channel-rate to determine an informative population-sampling procedure. We find that the algorithm can identify genotype superiority faster than the Moran model but at the cost of larger fluctuations in uncertainty. 
\begin{description}
\item[Keywords]
evolutionary theory, fitness landscapes, information theory
\end{description}
\end{abstract}
\maketitle

\section{Introduction}
One of the chief legacies of the modern evolutionary synthesis -- a period when evolutionary biology was modernized by mathematical and statistical formalisms---was the fitness landscape analogy, whereby genotype-phenotype space was likened to a multidimensional surface \cite{Wright, provine1989sewall,wade1992sewall}. It is one of the most popular subtopics of evolutionary biology, and has since been applied across a wide number of fields and problem types, ranging from antimicrobial resistance \cite{OgbunugaforEppstein2016}, to manufacturing \cite{mccarthy2000manufacturing}, clinical trials \cite{eppstein2012searching}, and many others. Such vast fitness landscapes are comprised of high-dimensional topographical features, with peaks and valleys characterized by ``ruggedness" \cite{Ruggedness} and notable network structures \cite{Aguirre, Yubero, Gavrilets}. These topographical features, crafted by forces like epistasis (the nonlinear interaction between mutations) \cite{epistasis,ogbunugafor2022mutation}, can dictate the direction \cite{weinreich2006darwinian} and speed \cite{ogbunugafor2019genetic} of evolution.

%As the environment changes (the standard for natural systems), so will the fitness values associated with the various genotypes. 

%Consequently, 

%The topography of the fitness landscape be defined and described by several features and metrics. These include the role of nonlinear interactions between mutations in producing "ruggedness" in in the topography \cite{Ruggedness}. More generally, studies on the topography of the fitness landscape are large in number, most discussing specifically how topography influences the direction \cite{weinreich2006darwinian} and speed of  \cite{ogbunugafor2019genetic} of molecular evolution. 

%Consequently, genetic variation can be defined with respect to how the variants (genotypes) vary with respect to traits such as longevity (a component of survival) or fecundity (how much they reproduce). 

%When a genotype has a relatively high reproductive fitness, it then represents a higher fraction of the population, whereas genetic variants with, and can even rise to "fixation," whereby a higher-fitness variant comes to dominate a population. 

%The collection of all genetic variants along with their associated fitness values is often referred to as a \textit{fitness landscape} in evolutionary biology \cite{Wright}. 

%Epistasis in particular represents an area of active investigation in fields such as evolutionary systems biology and population genetics today \cite{Weinreich2013a, epistasis ogbunugafor2022mutation}. 

A nuanced understanding of the topography of fitness landscapes is thought to be a key component to the potential for (i) predicting evolution at the molecular level and (ii) ``steering" evolution towards desired outcomes in fields like agriculture or biomedicine \cite{nichol2015steering}. For a current example: if the fitness landscape of an emerging virus could be understood in detail, it may be possible to predict which genotypes (or variants, in the context of SARS-CoV-2) are likely to arise. Other more classical examples include predicting the evolution of antimicrobial resistance \cite{PalmerKishony2013}, knowledge that could improve our ability to manage and treat infections. All of these problems are limited by theoretical and practical barriers to understanding fitness landscapes. 

The process of evolution is a natural mechanism for exploring these vast domains, especially as a means to locate relative peaks in a landscape. This natural algorithm has inspired the vast field of \textit{Evolutionary Algorithms} \cite{Crepinsek,Sloss2020}, of which \textit{Genetic Algorithms} represent a particularly popular model for exploration \cite{Holland1,Holland2}. In contrast, the question of how and to what extent the process of evolution itself should be regarded as a learning algorithm is scarcely considered. Our approach will be to view natural selection and drift as a kind of greedy algorithm for exploring and exploiting a fitness landscape, to borrow language from reinforcement learning \cite{Crepinsek}. In doing so, we will evaluate the efficiency with which this learning algorithm extracts information about the fitness landscape via changes in the population.

We note that the definition of information can be varied. In this work we concentrate on
determining which of two known landscapes is the current landscape, an idea which can easily be
extended to a palette of potential {\em a priori} landscapes. One can also imagine
information as the explicit genome $\rightarrow$ fitness map specification. This too would be
accessible by observation of a population roaming the landscape.  
Here we develop learning algorithms that explicitly maximize information extraction and we compare these to the efficiency with which evolution -- in this case the Moran model -- extracts the same information.  Evolution (under the Moran proxy) produces landscape-dependent population changes based on the
current population distribution. In contrast, our algorithms use a sequence of population
distributions as probes and measure the resulting birth/death responses as output.
These competing algorithms utilize a Bayesian approach that leverages tools from information
theory in an attempt to learn the landscape in an efficient manner. As
anticipated, these algorithms outperform the Moran evolution proxy in identifying which landscape is in
use. We present these algorithms primarily to provide a theoretical point of reference with which to compare to the efficiency of evolution, regarded as an information-extraction mechanism. Nevertheless, these algorithms could in principle be adapted to laboratory settings to efficiently characterize particular features of a landscape.

The algorithm works by quantifying the information communicated to an evolving population through
the selective pressures of an environment, namely the information associated to which genotype has superior fitness. The selective pressures of an environment tend to shift the population
distribution in favor of a particular genotype. It is in this sense that we say ``evolution learns
a landscape": The population of variants is redistributed in accord with the fitness landscape. We
analogize this exchange between an environment, which determines the fitness landscape, and the
population distribution as a communication channel, where the particulars of the landscape are
communicated to the population in the form of population-shifts, $\Delta$. The associated
``channel-rate" quantifies the flow of information that is communicated over the channel. We will use this channel-rate to establish when population shifts can be most informative in identifying the superior genotype.

Prior work has formulated similar analogies between evolving populations and communication channels \cite{Gong, Watkins}. In the case of \cite{Gong}, it is the genetic translation of proteins from one generation to the next that is regarded as a noisy channel, owing to the error-prone nature of transcription and translation. In \cite{Watkins}, the author uses the framework of selective breeding to quantify the extent to which an organism's \textit{genome} encodes the selective pressures of its environment. In that framework, a channel capacity for sexual and asexual breeding is computed and compared. 

Our work differs from \cite{Gong} in that we abstract away the details of the evolutionary process to focus on a general context in which evolutionary pressures are encoded in a collection of organisms, which is more akin to \cite{Watkins}. In contrast to \cite{Watkins}, however, our work considers the extent to which evolutionary pressures are encoded in the \textit{population} distribution of genotypes, or more specifically, in the changes of the population distribution, rather than in the genome itself.

In section II we provide a brief primer on the Moran model of evolution. In section III we formulate a communication channel analogy and compute its channel-rate as a function of the population distribution. In section IV, we introduce our algorithm and present a comparison to the Moran model, viewed as a competing learning algorithm. We conclude in section V with a discussion of some implications of this comparison and suggest future directions.

\section{A discrete model of selection and drift}
\label{sect:discretemodel}
The Moran process \cite{Moran} is a commonly used Markov model in evolutionary biology to allow evolution to play out in discrete time on a fixed population of size $N$. In the two-genotype case we can characterize the fitness landscape with the ratio of fitness values,
\be
\label{eq:sdef}
    r = \frac{f_A}{f_B}
\ee
where $f_A, f_B \in (0,1)$ measure ``fitness" of variants $A$ and $B$ in a given environment. In the following model, only the ratio of fitness values will figure into the evolutionary dynamics and so the space of fitness landscapes, which we call $\Fspace(r)$, can be parameterized by $r \in (0,\infty)$, the \textit{relative fitness} of the $A$-type.  

In particular, the relative fitness determines the birth rate of $A$-type individuals in the population. We define $\phi_A(r,k)$ to be the probability that an $A$-type produces an offspring in a given (discrete) time step of the model, where $k\in\{0,1,...,N\}$ is the population size of the $A$ variant. Specifically, we set
\be
\label{eq:phidef}
\phi_A(r,k) \equiv \frac{f_A k}{f_A k + f_B (N-k)} = \frac{rk}{(r-1)k+N}
\ee
In a landscape where $r > 1$, the $A$-type has the reproductive advantage, whereas a corresponding landscape with
fitness ratio $1/r$ offers that same advantage to the $B$-type instead. More generally, in cases with $n\ge 2$
genotypes, an $n-1$ length vector of $r$-values would be necessary to
characterize the landscape. Likewise, in the $n=2$ case, the population space $\Pspace$ can be characterized by just the $A$-population $k$ -- since the $B$-population is $N-k$, and $N$ will be held fixed by the dynamics -- whereas an $n-1$ length vector of $k$-values would be necessary
for models with $n$ genotypes. As we outline below, the relative fitness {\em only}
determines the reproductive rates, whereas death rates are strictly proportional to the population
values. Although models where fitness affects death have been considered as well
\cite{viability-selection, death-birth}.

The Moran model holds the population constant through the following 3-step process:
\begin{itemize}
\item  {\bf Birth:}  $X \to X + \Xd$
  \begin{itemize}
  \item
    A genotype is selected to produce a daughter, $\Xd$: $A$-type selected with probability $\phi_A$ and $B$-type with $1-\phi_A$.
  \end{itemize}
\item  {\bf Death:}  $X \to \mbox{\O}$ 
  \begin{itemize}
  \item
    Exactly one individual (other than $X_d$) is uniformly selected for removal.
  \end{itemize}
\item  {\bf Replacement:}  $\Xd \to X$
  \begin{itemize}
  \item
    The daughter replaces the decedent.
  \end{itemize}
\end{itemize}
That is, we assume a population of $N$ individuals. Birth occurs asexually with a member of a given
variant selected for reproduction with a probability that scales with the fitness of that variant, $f_A$
or $f_B$, as in equation~\ref{eq:phidef}.  Death occurs by randomly selecting one of the $N$ population members {\em not}
including the daughter and deleting it. The daughter then replaces the decedent, keeping the population constant. We refer to this 3-step process as a single \textit{Moran step}. 

Mutation is also often included in the Moran model by incorporating a step where the genotype can switch to the opposite type with some fixed probability -- we do not consider this here since the mutation probability is usually low ($p_{mutation}\ll1$) \cite{Moran}, and our analysis can be greatly simplified without it. A consequence of setting $p_{mutation}=0$ will be that the $k=0$ and $k=N$ population values are absorbing boundaries in our model.

This seemingly odd death mechanism (in that death is independent of fitness), is a hallmark of the
well-studied Moran process and captures the essence of a population with a limited resource that is
sufficiently well-mixed so that a single birth can only be accommodated by a single death, on
average \cite{Moran, spatial-Moran, amplified-Moran, death-birth}. Our model thus ignores any
spatial structure that may be tied to resource-allocation, although in some models death occurs in
the proximity of births \cite{spatial-Moran} or is tied to fitness \cite{death-birth}.

\begin{figure}[ht]
    \centering
%    \vspace{-0.15in}
    \includegraphics[width=3.5in]{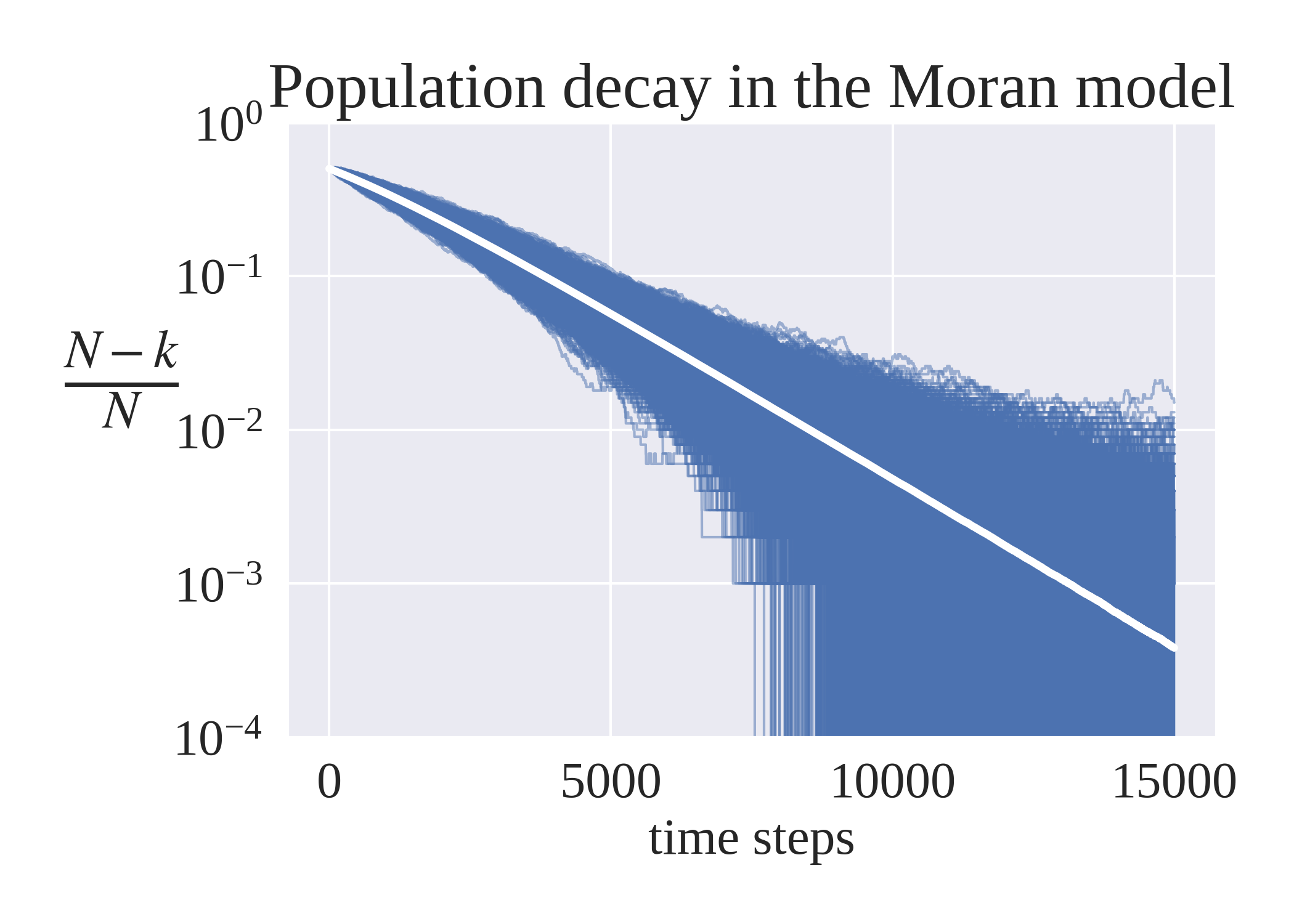}
%\vspace{-0.4in}
    \caption{{\bf Moran Model Example Simulation:} We show a sample of $6,000$ (half of total) population trajectories for $r=2$, $N=1000$. We initialize the $A$ population fraction to $1/2$. The superimposed curve shows the mean of $12,000$ simulation runs.}
    \label{fig:simulations}
\end{figure}

An example evolution time course is provided in FIGURE~\ref{fig:simulations} for $N=1000$ and $r=2$, which favors the $A$ population. We show how the $B$-type population fraction $(N-k)/N$ decays to zero as the $A$ population takes over. We run $12,000$ random independent trials and plot half of these trajectories (due to rendering difficulties). The average of all $12,000$ trials is superimposed in white. Later we interpret the population fractions as a measure of the confidence the evolving system has in one genotype's superiority over the other. We will see that a corresponding measure of confidence by our Bayesian algorithm can converge several times faster. In this respect, we will argue that evolution is relatively slow in learning the landscape.

%\vspace{-0.017in}

\section{What Does $\Delta_k$ Say About $\Fspace$}
\label{sect:dchannel}

In this work we regard the Moran model as an algorithm for learning a landscape. In the
related context of Genetic Algorithms for example, one regards members of a population as
``strategies" run in parallel to determine which yields superior performance \cite{Holland1,Holland2}. We will likewise
regard the fraction of genotypes in the Moran process as a proxy for the population's confidence about genotype superiority in that landscape. For our purposes, we will only consider two landscapes,
$\Fspace_A=\Fspace(r_A)$ and $\Fspace_B=\Fspace(r_B)$, one where the $A$-type is dominant ($r_A>1$) and one where the $B$-type is dominant in a symmetric way; i.e. $r_B=1/r_A<1$. Thus, distinguishing between landscapes will be equivalent to determining genotype superiority. In this way, we will be able to form a comparison between the confidence that
the Moran model develops, by way of updating the population distribution, and our own competing
Bayesian algorithm, by way of updating priors over the space of landscapes.

To be clear, our approach is to ask how
population changes can be used to infer the current landscape, $\Fspace_A$ or $\Fspace_B$. In related literature, however, the focus has primarily been on
whether it is the $A$-type or the $B$-type that has superior fitness in a given environment. After
all, in a static environment the crucial information is which variant has optimal fitness in order
that one's ``bet" on a variant would minimize loss, usually measured by some form of ``regret"
\cite{McGee} or ``substitutional load" in the context of evolutionary biology \cite{Kimura}. We bridge this approach and ours by restricting ourselves to the case that $r_A = 1/r_B$ so that identifying the landscape is equivalent to identifying genotype superiority. 

Define $\Delta_{k_t}=k_{t+1}-k_t\in\{-1,0,1\}$ to be the change in the $A$
population, $k_t$, after a Moran step. Our goal will be to determine which of the two possible landscapes
$\{\Fspace_A,\Fspace_B\}$ is in use -- only one landscape will be in use throughout each run, which we arbitrarily take to be $\Fspace_A$. Essentially, we would like to
distinguish between two Markov models that differ by their set of transition probabilities,
$p(k+\Delta_k|k)$, and to do so quickly. 

\begin{figure}[ht]
    \centering
    \includegraphics[width=2.5in]{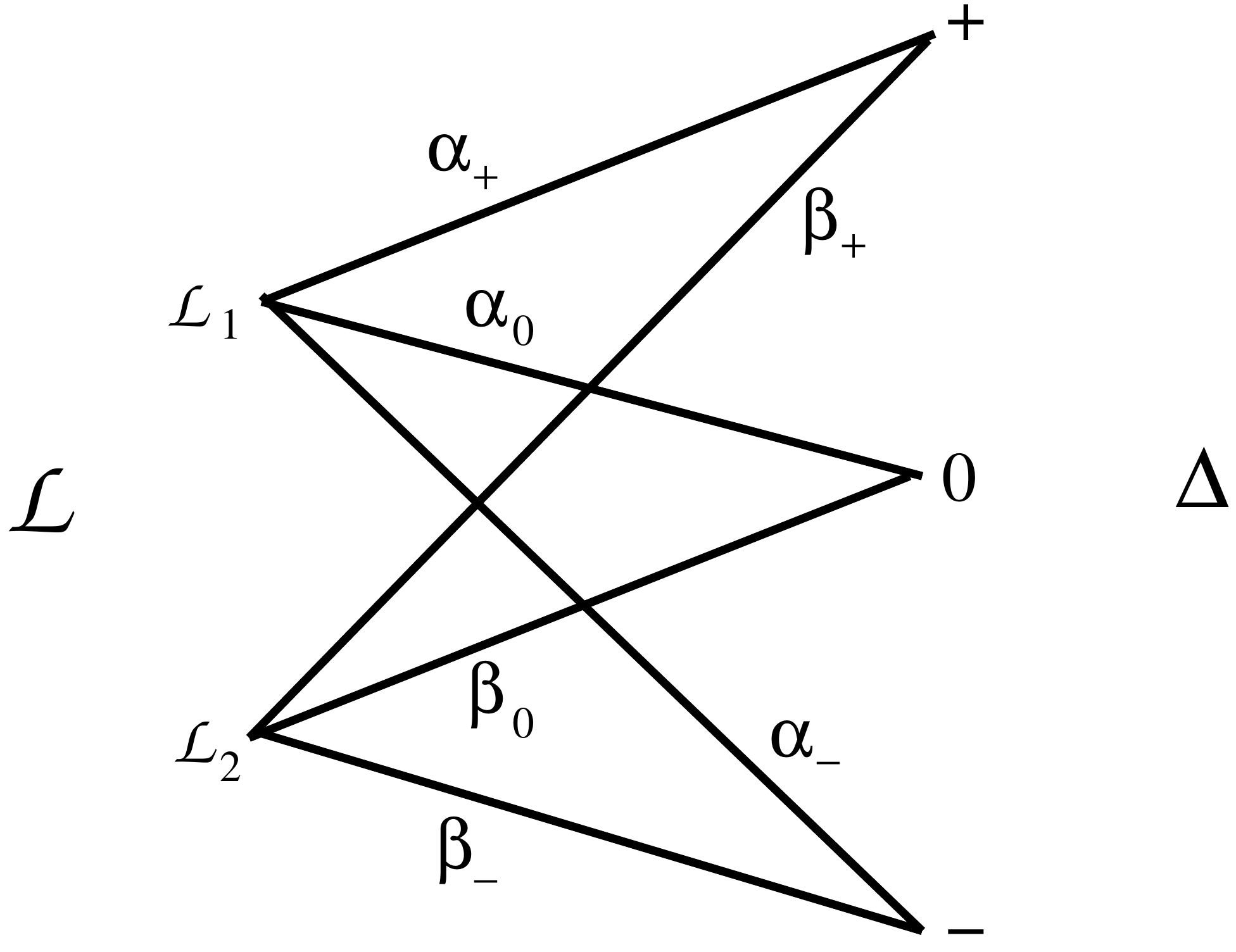}
%\vspace{-0.1in}
    \caption{{\bf \boldmath The $\Fspace \to \Delta$ Channel:} transition probabilities $\alpha_\Delta$ and $\beta_\Delta$ as defined in the text. }
    \label{fig:deltacollapsed}
\end{figure}

We abstract this problem to a communication channel, shown in FIGURE~\ref{fig:deltacollapsed}, between the space of possible landscapes, $\{\Fspace_A,\Fspace_B\}$, and the space of population transitions, $\Delta_k \in\{-1,0,1\}$, originating at population $k$. Our channel analogy captures the sense that the environment (landscape) is communicated to the population through selective pressure on the population. Strictly speaking, we have a collection of channels indexed by the population $k$ -- the transition probabilities of FIGURE~\ref{fig:deltacollapsed} are functions of $k$ and $r$ -- so our model is actually more analogous to a \textit{finite state Markov channel} (FSMC), used to model fading in communication channels \cite{Wang}. 

Changes in the population, $\Delta_k$, encode information about genotype superiority. The maximum amount of information is quantified by the
channel-rate, which is given by the mutual information $I(\Delta_k;\Fspace)$. For a given value of $k$, $I(\Delta_k;\Fspace)$ is a measure of how informative the transition from $k$ to $k+\Delta_k$ is in distinguishing between $\Fspace_A$ and $\Fspace_B$. 

We label the transition probabilities $p(k+\Delta_k|k)$ for the two models by $\alpha_\Delta$ and
$\beta_\Delta$ (suppressing their $k$-dependence for notational convenience), which depend on $r_A$
and $r_B$ respectively. The explicit forms of $\alpha_\Delta$ and $\beta_\Delta$ are given 
 below (we use $+$ and $-$ as shorthand for $+1$ and $-1$
respectively):
\begin{align*}
    \alpha_+ &=  \phi_A(r_A,k)\frac{N-k}{N}=\frac{f^{(\Fspace_A)}_Ak(N-k)}{\left(f^{(\Fspace_A)}_Ak+f^{(\Fspace_A)}_B(N-k)\right)N}, \\
    \alpha_- &=  (1-\phi_A(r_A,k))\frac{k}{N}=\frac{f^{(\Fspace_A)}_Bk(N-k)}{\left(f^{(\Fspace_A)}_Ak+f^{(\Fspace_A)}_B(N-k)\right)N}, \\
    \alpha_0 &= 1 - \alpha_{+} - \alpha_{-} = \frac{f^{(\Fspace_A)}_Ak^2 + f^{(\Fspace_A)}_B(N-k)^2}{\left(f^{(\Fspace_A)}_Ak+f^{(\Fspace_A)}_B(N-k)\right)N}
\end{align*}
where $f^{(l)}_A$ and $f^{(l)}_B$ are the fitness values of the $A/B$ types in landscape $l\in\{\Fspace_A,\Fspace_B\}$. $\beta_+,\beta_-$ and $\beta_0$ take the exact form as above for $\alpha_+,\alpha_-$ and $\alpha_0$, respectively, but with the ratio $r_A=f^{(\Fspace_A)}_A/f^{(\Fspace_A)}_B$ replaced by its reciprocal, $r_B=f^{(\Fspace_B)}_A/f^{(\Fspace_B)}_B=f^{(\Fspace_A)}_B/f^{(\Fspace_A)}_A$.
Note that when $r_A = r_B$ ($\Fspace_A$ and $\Fspace_B$ identical), then ${\alpha_\Delta}={\beta_\Delta}$. In which case, the channel-rate is identically zero.

\begin{figure}[ht]
    \centering
%    \vspace{-0.2in}
    \includegraphics[width=3.5in,height=2.5in]{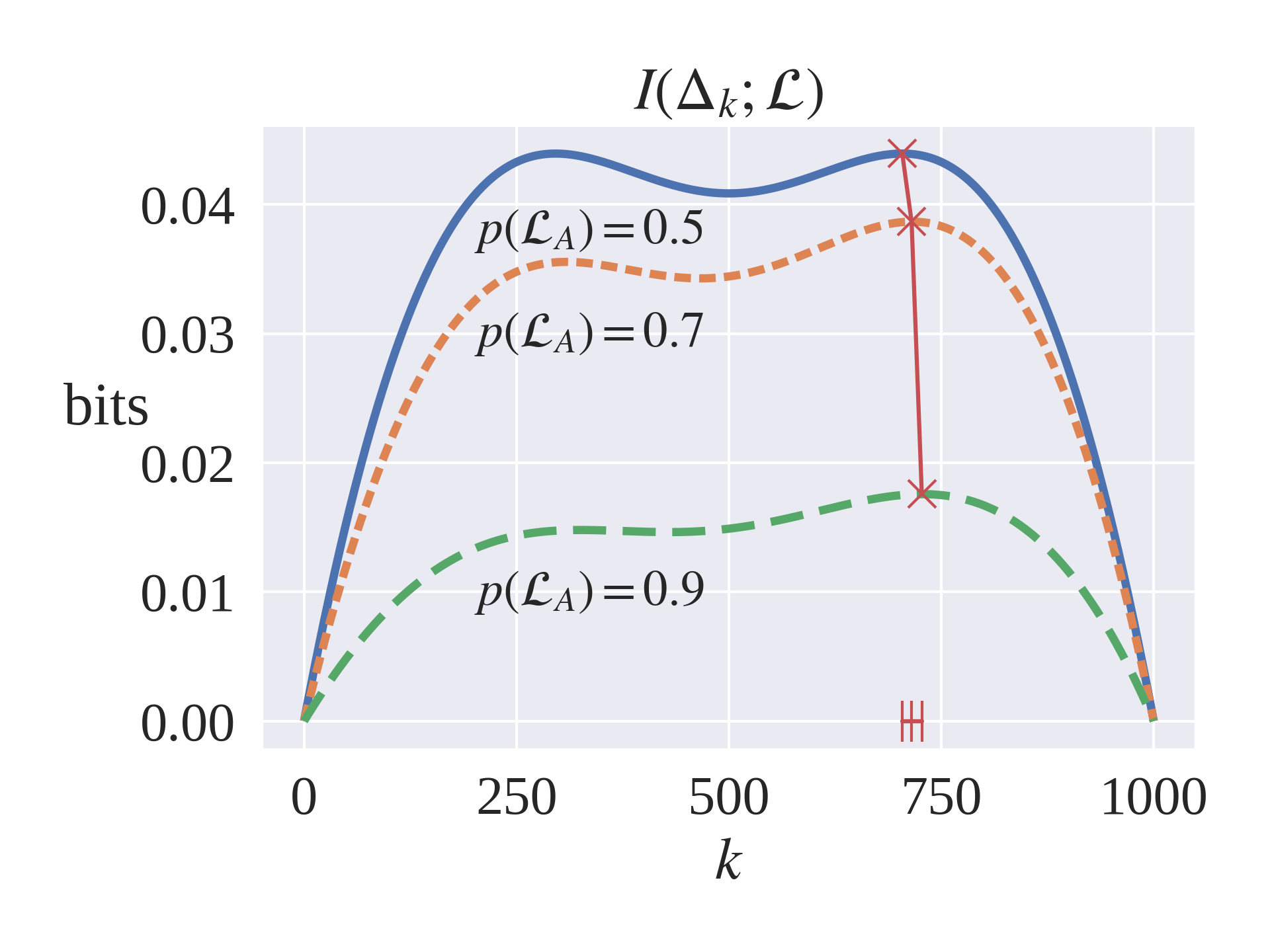}
%    \vspace{-0.5in}
    \caption{{\bf \boldmath $\Dchannel$ Rate, $I(\Delta_k, \Fspace)$:} We set $r = r_A = 1/r_B = 2$, we consider $p(\Fspace_A)\in\{0.5,0.7,0.9\}$ and label $\max(I)$ and the corresponding population value.}
    \label{fig:deltaRate}
\end{figure}

In FIGURE~\ref{fig:deltaRate}, we compute $I(\Delta_k;\Fspace)$, setting $r = r_A = 1/r_B = 2$
and considering several values for the prior over the $\Fspace_A$ landscape: $p(\Fspace_A) \in  \{ 0.5,\:0.7,\:0.9\}$  (the $\Fspace_B$ prior is $1$ minus the $\Fspace_A$ prior). These curves show where population changes are maximally informative and can be used to quickly distinguish between landscapes, as we see in the next section. 

%\vspace{-0.1in}

\section{Bayesian inference vs. Moran}
\subsection{An aside on Replicator Dynamics} 
Before discussing our algorithm we should briefly mention another popular model of evolution, \textit{Replicator Dynamics}. This models selective pressure in a deterministic way and has been formally analogized to Bayesian learning \cite{Harper2,Czegel1,Czegel2,Watson}. The analogy is formed by regarding the genotype population fractions as
priors over the space of genotype superiority, and regarding updates to the population distribution as
analogous to Bayesian updates on the priors \cite{Harper2}. In the \textit{discrete} Replicator Dynamics, the update rule precisely mimics Bayesian updating \cite{Harper1}. Whereas in the \textit{continuous} case, the differential equation describing the population dynamics, which is a weak selection ($r \approx 1$) and large population ($N \gg 1$) limit of the Moran model, is that of a gradient flow, increasing the population's average fitness \cite{Harper1,Hofbauer}. 

Both the discrete and the continuous models increase the average fitness of the population by shifting the population toward the
superior genotype \cite{Harper1}. This procedure however is limited to making local steps in the space of population distributions, and thus local steps in inference with regards to genotype superiorty. Furthermore by virtue of being a gradient that prioritizes fitness ascension, the continuous Replicator Dynamics tends to flow away from more informative points in
the population distribution space, moving toward suboptimal regions for the purposes of identifying genotype superiority. In particular, as the population flows along the
fitness gradient, the population distribution moves away from the point of peak
mutual information between the space of population shifts and the space of superior genotypes. This is
understandable since in addition to being an exploration algorithm, evolution is simultaneously an
exploitative algorithm that greedily moves towards improved fitness. The same can be said about the
Moran process, as we will see.

Our approach differs from the analysis of Replicator Dynamics in two notable forms: (i) the Moran process is a stochastic model that can be used to interrogate finer-scale
fluctuations in the population, and (ii) we
position our sense of optimality on the efficiency with which information can be extracted from the
environment, rather than by improving the population fitness rapidly through local steps in the population distribution. The Moran model also has not, to our knowledge, been considered in a learning context

Analogous to Replicator Dynamics, however, we will regard the Moran model as a learning algorithm that uses the population distribution as a measure of confidence in genotype superiority. In this \textit{Moran algorithm}, the population is updated according to natural selection. Whereas for our competing algorithm we form an explicit prior over the landscapes as a measure of the confidence in genotype superiority, and we update the prior according to Bayes' rule. We discuss our algorithm next.

\subsection{A Bayesian learning algorithm}
From our calculation of $I(\Delta_k;\Fspace)$, we see that we can identify the population value $k^*$ from which population changes are most informative. Our Bayesian
learning algorithm runs a Moran step from the population value $k^*$, and uses the population shift $\Delta_{k^*}$ to update a prior over $\Fspace_A$ at step $t$, which we now denote by $p_t=p(\Fspace_A)$ (and we set $p_0=1/2$). Note that earlier we used $p(\Fspace_A)$ to generically denote the prior. However, now we consider a sequence of priors $(p_0,p_1,...,p_t)$ and hence adopt a time-dependent notation. We will also write $\Fspace^t\in\{\Fspace_A,\Fspace_B\}$ to emphasize that the landscape has a time-dependent prior (even though the underlying landscape itself is assumed to be \textit{static}). Our algorithm then picks the landscape with the maximum prior: $p_t$ for $\Fspace_A$ or $1-p_t$ for $\Fspace_B$. We summarize our algorithm below:

\underline{``$k=\argmax(I)$" algorithm}:
\begin{enumerate}
     \item[(1)] Compute $I(\Delta_k;\Fspace^t)$ using $p_t$ for each $k$
    \item[(2)] Find $k^*=\argmax_k I(\Delta_k;\Fspace^t)$
    \item[(3)] Run Moran step to generate $\Delta_{k^*}$
    \item[(4)] Use $\Delta_{k^*}$ to compute $p_{t+1}$ according to Bayes' rule
    \item[(5)] Declare $\Fspace_A$ if $p_{t+1}>0.5$, else declare $\Fspace_B$
\end{enumerate}
$p_t$ is updated according to Bayes' rule in the following manner. $\Delta_{k^*}$ is generated by a Moran step and we compute the transition probability $p(\Delta_{k^*}|\Fspace_A)$, which gives the probability for a population change $\Delta_{k^*}$ starting from $A$-population $k^*$, when landscape $\Fspace_A$ is in use. For a given value of the prior $p(\Fspace_A)$ and population change $\Delta_k$, we can use the transition probability $p(\Delta_{k^*}|\Fspace_A)$ to compute the conditional probability $p(\Fspace_A|\Delta_k)$ according to Bayes' formula,
\be
p(\Fspace_A|\Delta_k)=\frac{p(\Fspace_A)p(\Delta_k|\Fspace_A)}{p(\Fspace_A)p(\Delta_k|\Fspace_A)+p(\Fspace_B)p(\Delta_k|\Fspace_B)}
\ee
Using $p_t$ to refer to $p(\Fspace_A)$ at a particular time step (and $1-p_t$ for $p(\Fspace_B)$) and setting $p_{t+1}=p(\Fspace_A|\Delta_k)$ to be the updated prior, Bayes' rule can be written as, 
\be
p_{t+1} = \frac{p_t}{p_t + (1-p_t)\gamma_t} = \frac{1}{1+\left(\frac{1-p_t}{p_t}\right)\gamma_t}
\label{eq:p_t}
\ee
where we have divided out the numerator and denominator by $p(\Delta_k|\Fspace_A)$, and set $\gamma_t$ to be,
\be
\gamma_t = \frac{p(\Delta_{k_t}|\Fspace_B)}{p(\Delta_{k_t}|\Fspace_A)}
\ee
A considerably simpler form can be written in terms of the likelihood ratio, $q_t=(1-p_t)/p_t$. Under this substitution, equation~\ref{eq:p_t} can be written, 
\be
q_{t+1} = q_t \gamma_t
\label{eq:LRT}
\ee
from which we see that,
\be
q_t = \gamma_{t-1}\gamma_{t-2}...\gamma_0
\label{eq:gamma_product}
\ee
($q_0=1$ since $p_0=1/2$). Note that we expect $q_t\rightarrow 0$ ($p_t\rightarrow 1$) when landscape $\Fspace_A$ is in use. In particular, using equation~\ref{eq:gamma_product} we can write $p_t$ as, 
\be
p_t=\frac{1}{1+\gamma_{t-1}...\gamma_0}=\frac{1}{1+e^{\sum_i\ln(\gamma_{i})}}
\ee
and $\ln(\gamma_i)=\ln p(\Delta_{k_i}|\Fspace_B)-\ln p(\Delta_{k_i}|\Fspace_A)$, which will be negative on average when landscape $\Fspace_A$ is in use. Thus $p_t$ is expected to tend to $1$. 

Note that $q_{t}\gamma_t=\frac{1-p_t}{p_t}\frac{p(\Delta_{k_t}|\Fspace_B)}{p(\Delta_{k_t}|\Fspace_A)}$ takes the form of a likelihood ratio test. In fact our decision rule picks landscape $\Fspace_A$ when $p_{t+1}>0.5$, which is equivalent to picking $\Fspace_A$ when $q_{t+1}<1$. And so our ``$\argmax(I)$" algorithm is equivalent to a likelihood ratio test with a Bayesian-updating prior. 

\subsection{Convergence rates}
We can compute how quickly the Moran algorithm and our Bayesian approach will reduce their uncertainty. In the Moran case, the average population distribution $\kappa(t) = \langle k_t \rangle/N$ (averaged over an ensemble of trials) satisfies a logistic equation of the form,

\be
\frac{d \kappa}{dt} = \left(\frac{r-1}{N}\right)\kappa(\kappa-1).
\label{eq:logistic}
\ee

When $\Fspace_A$ is in use, $\kappa(t)\rightarrow 1$ as $t\rightarrow \infty$ (or at least for $t\gg N/(r-1)$) and so if we set $\kappa(t) = 1-\epsilon(t)$ for $\epsilon(t)\ll 1$, then $\epsilon(t)$ satisfies,

\be
\frac{d\epsilon}{dt} = - \left(\frac{r-1}{N}\right) \epsilon + O(\epsilon^2).
\ee 
For $\epsilon$ sufficiently small, the solution will take the form: $\epsilon(t)\approx\epsilon(0)\exp(-\frac{r-1}{N}t)$, and thus the solution $\kappa(t)\approx 1-\epsilon(0)\exp(-\frac{r-1}{N}t)$ decays to unity at rate $\frac{r-1}{N}$. In particular, convergence is slowed by large population sizes $N$, and accelerated for landscapes with large selective pressure ($r\gg 1$).

In the Bayesian case, we have seen that $p_t$ takes the form $p_t=\left(1+\exp\sum_i \ln(\gamma_i)\right)^{-1}$. As the algorithm develops confidence in the $\Fspace_A$ landscape (our default underlying landscape), $p_t$ approaches $1$, and in that limit our choice of the probe population $k^*$ settles on a particular value $k_\infty^*$ (for $N=1000$ and $r=2$, the probe population settles on $k_\infty^*=734$, for instance). In that limit, the population steps are sampled i.i.d. as $\Delta_{k_\infty^*}\sim\{p(+1|\Fspace_A),p(-1|\Fspace_A),p(0|\Fspace_A)\}$, where it is understood that $p(\delta|\Fspace_A)$ depends implicitly on $k_\infty^*$. As more time accumulates with $k^*=k_\infty^*$, the value $\frac{1}{t}\sum_{i=1}^t \ln(\gamma_i)$ approaches its average $\overline{\ln(\gamma)} = \sum_\delta p(\delta|\Fspace_A) \ln(\gamma_\delta)$, with $\delta\in\{+1,-1,0\}$ -- we abuse notation somewhat with $\gamma_\delta\equiv p(\delta|\Fspace_B)/p(\delta|\Fspace_A)$. This average is nothing more than the negative Kullback-Leibler divergence between the conditional distributions: $\overline{\ln(\gamma)}=-D(A||B) = -D\left(p(\Delta_{k_\infty^*}|\Fspace_A)||p(\Delta_{k_\infty^*}|\Fspace_B)\right)\le0$ \cite{Cover}. Thus as $t\rightarrow\infty$ we have that,
\be
p_t \rightarrow \frac{1}{1+e^{-tD(A||B)}}.
\ee
This shows that for large $t$ ($\gg 1/D(A||B)$), the convergence rate of $p_t$ is limited only by how distinguishable population changes $\Delta_k$ are between $\Fspace_A$ and $\Fspace_B$. This is to be contrasted with the Moran case, which is limited by the population size $N$ and the selection strength $r$. 

\subsection{Comparing learning algorithms}
\subsubsection{``pick $k$" algorithms}
The Bayesian procedure outlined above selects population values $k^*$ from which to run episodes of a maximally
informative nature. We can compare this approach to other, more naive, ``pick $k$" algorithms. One
such approach is to instead pick $k$ uniformly randomly among $\{0,1,...,N\}$ in step (2), and
otherwise retaining steps (3) -- (5). 

Another approach picks $k$ according to the walk that $k$ takes in the Markov chain, induced by
the Moran process -- what we call a ``Moran walk." In that algorithm, a stochastic sequence of
$k$-values is provided by the Markov model itself: $(k_0,k_1,...,k_t)$. At each step
$\Delta_{k_t}=k_{t+1}-k_t$ is generated and steps (4) -- (5) can be performed in the same manner as
above. This is an online type of algorithm and compares rather directly to the Moran process, which
viewed as an online algorithm itself uses the population distribution, $(k/N,(N-k)/N)$, to distinguish between landscapes, rather than some external prior like our Bayesian approaches. For the Moran process, we take the view that the majority population is an attempt by natural selection to ``decide" on genotype superiority. And so we refer to the Moran algorithm as the process of deciding on the landscape using the $50/50$ mark of the population. So that if $k/N>0.5$, for example, then $\Fspace_A$ is chosen at that Moran step. We summarize the essential features of each algorithm in TABLE~\ref{table:table1}.

\begin{table*}[ht]
\caption{Overview of algorithms}
\centering
% \begin{tabular}

\begin{ruledtabular}
\begin{tabular}{cccc}
Algorithm name & Bayesian & How is $k$ generated? &  Decision rule\\
\hline & \\[-7pt]
$\boldsymbol{\argmax(I)}$ & Yes & Maximizes $I(\Delta_k;\Fspace)$ & $p_t>0.5\rightarrow\Fspace_A$\\
\textbf{Random} $\boldsymbol{k}$ & Yes & Uniformly randomly & $p_t>0.5\rightarrow\Fspace_A$\\
\textbf{Moran walk} & Yes & According to Markov chain & $p_t>0.5\rightarrow\Fspace_A$\\
\textbf{Moran algorithm} & No & According to Markov chain & $k/N>0.5\rightarrow\Fspace_A$
\label{table:table1}
\end{tabular}
\end{ruledtabular}
\end{table*}

From this perspective, we regard the population distribution as a kind of probe of the fitness landscape, from which information can be extracted in the form of population changes. Our algorithm chooses a constantly-updating probe that is tuned to yield more information. Whereas, natural selection employs a probe which is also updated with each Moran step, but is not necessarily optimal for information extraction. 

\subsubsection{Empirical probability of error}
\begin{figure}[ht]
    \centering
    \includegraphics[width=3.5in,height=2.5in]{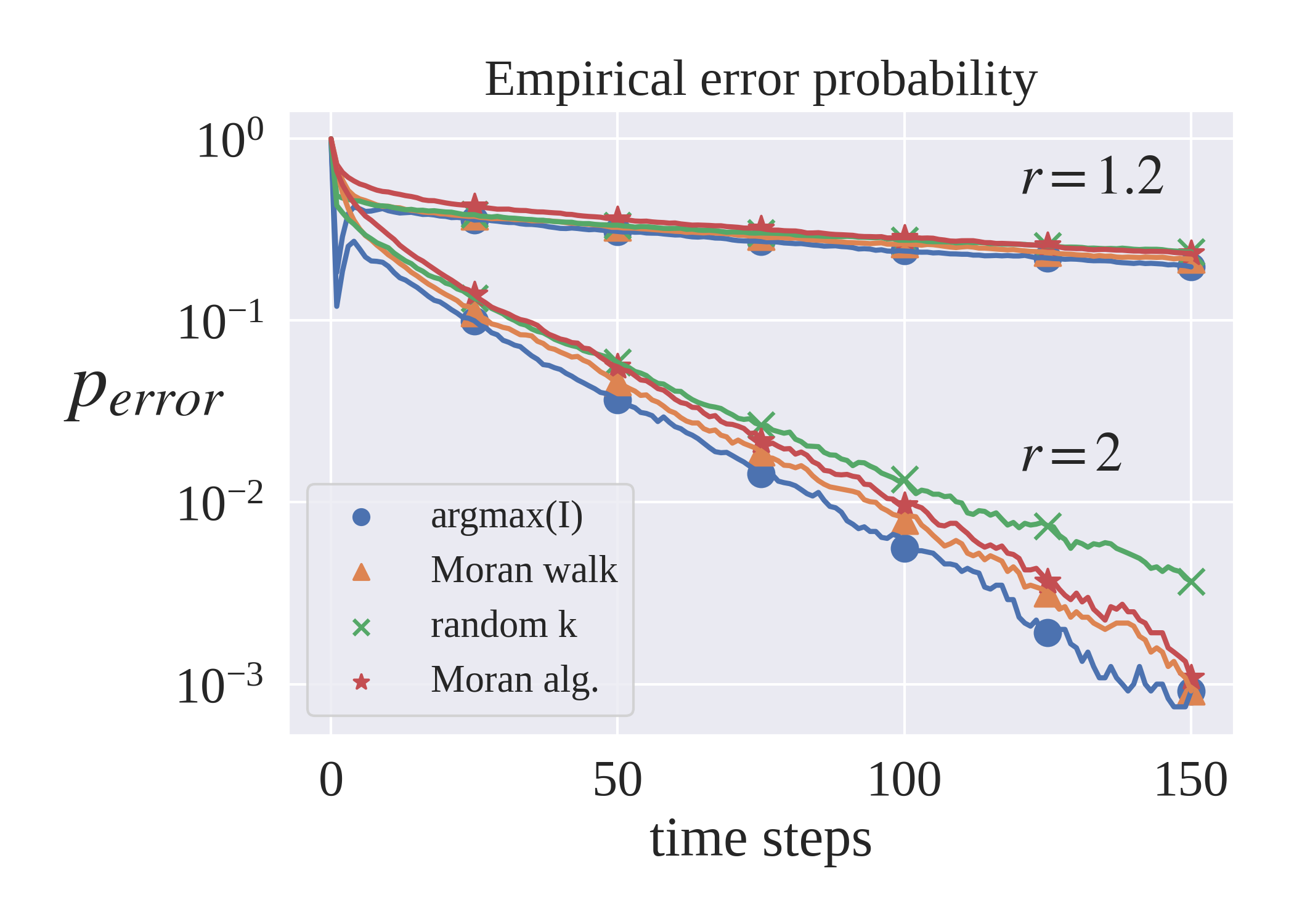}
%\vspace{-0.4in}
    \caption{{\bf \boldmath``empirical error probability" $p_e$: } True landscape set to $\Fspace_A=\Fspace(2)$, each curve shows the fraction of $12,000$ independent trials run for 150 episodes each where that algorithm incorrectly identifies the landscape. The mean of the $12,000$ trials is superimposed (white). The ``Moran walk" algorithm uses the sequence of $k$-values generated by the Moran algorithm in each trial.} \label{fig:p_e}
\end{figure}

We quantify the reduction of uncertainty about the true landscape with
an empirical probability of error for each algorithm, $p_{error}(t) = n_e(t)/\mathcal{N}$, where $n_e(t)$ counts the number of trials out of $\mathcal{N}$ where an algorithm declares the incorrect landscape at algorithmic step $t$ (in all cases the true landscape is set to $\Fspace_A$).
In FIGURE~\ref{fig:p_e} we show how the Moran model, treated as a learning algorithm in the way just described, compares to our algorithm. We also include the two other naive ``pick $k$" Bayesian approaches for comparison: ``random $k$" and our ``Moran walk." We set $r=2$ so that $\Fspace_A=\Fspace(2)$ and
$\Fspace_B=\Fspace(1/2)$, so that type $A$ has twice the fitness as
type $B$ in $\Fspace_A$, and type $B$ has twice the fitness of type $A$ in $\Fspace_B$. Also shown is the $r=1.2$ case for comparison. We show the results of $\mathcal{N}=12,000$ independent trials for each algorithm, each run for 150 steps. $p_t$ was initialized to 0.5 for each trial in the Bayesian approaches.

We find that our ``$k=\argmax(I)$" algorithm reduces the probability of error most rapidly. We also see that the Moran algorithm does perform better than the random Bayesian approach (``random $k$") and otherwise tracks well with the ``Moran walk" approach. The advantage of the $k=\argmax(I)$ approach seems to come from optimizing the population probe in testing the landscapes.

\begin{figure}[ht]
    \centering
    \includegraphics[width=3.5in,height=2.5in]{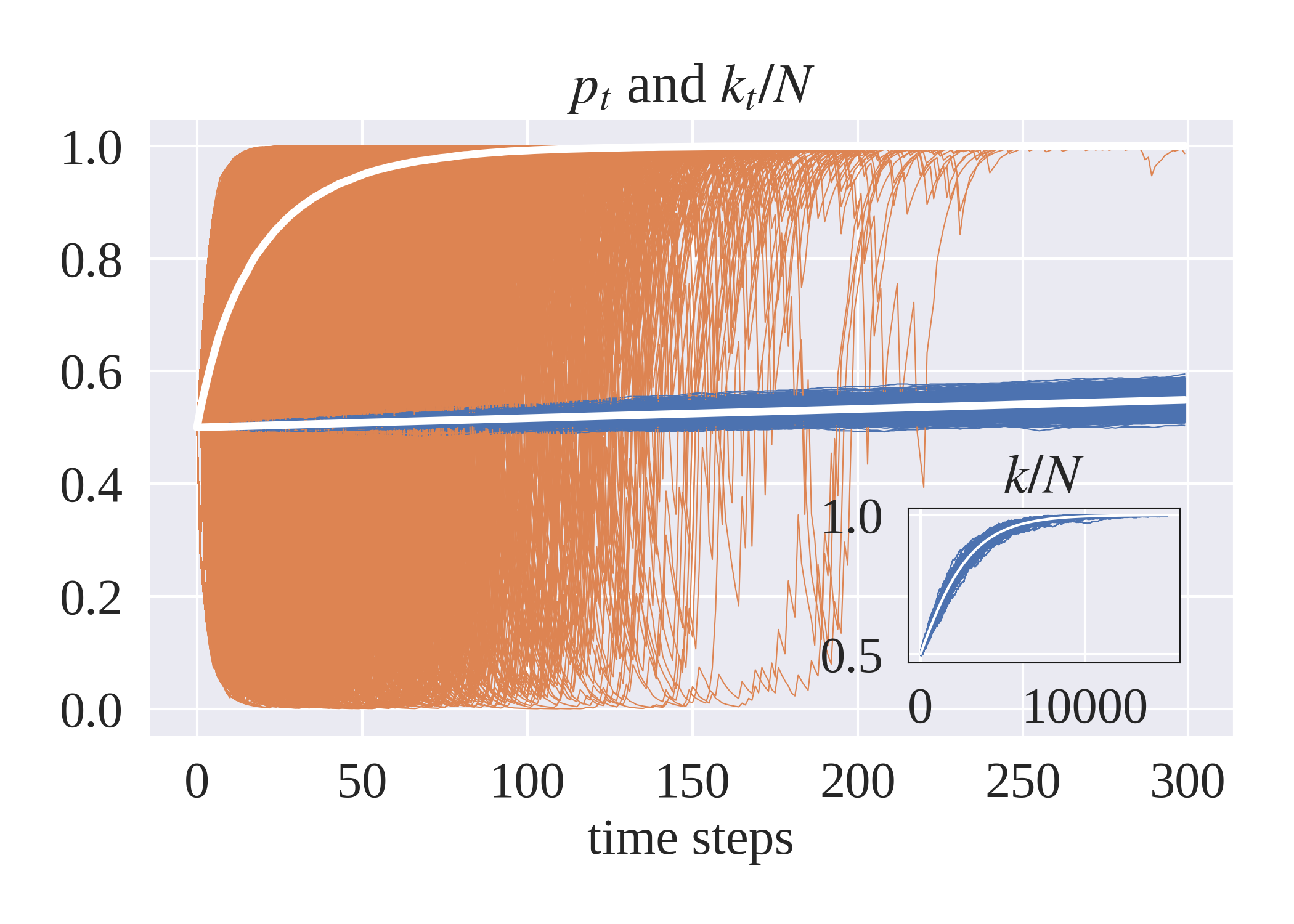}
%\vspace{-0.4in}
    \caption{{\bf \boldmath``$p_t$ and $k_t/N$": } We compare two proxies for certainty in genotype superiority: $p_t$, the prior over $\Fspace_A$ in our Bayesian algorithm (orange), and $k_t/N$ (blue), the $A$-type population fraction in the Moran algorithm. $12,000$ independent trajectories are shown in each case, with their average superimposed (white). The inset shows a zoomed out picture of $k/N$ (blue).} \label{fig:slow_and_steady}
\end{figure}

FIGURE~\ref{fig:slow_and_steady} directly compares our proxies for the algorithms' confidence in identifying the correct landscape (we show the $k=\argmax(I)$ algorithm and the Moran algorithm, for $r=2$). The average over $12,000$ $p_t$-trajectories in the $k=\argmax(I)$ case plateaus on the order of $10^2$ time steps. Whereas, the proxy in the Moran algorithm, the $A$-type population fraction $k/N$, plateaus by a comparable extent on the order of $10^4$ time steps (see FIGURE~\ref{fig:simulations} for detailed limiting behavior of $1-k/N$). 

\begin{figure}[ht]
    \centering
    \includegraphics[width=3.5in,height=2.5in]{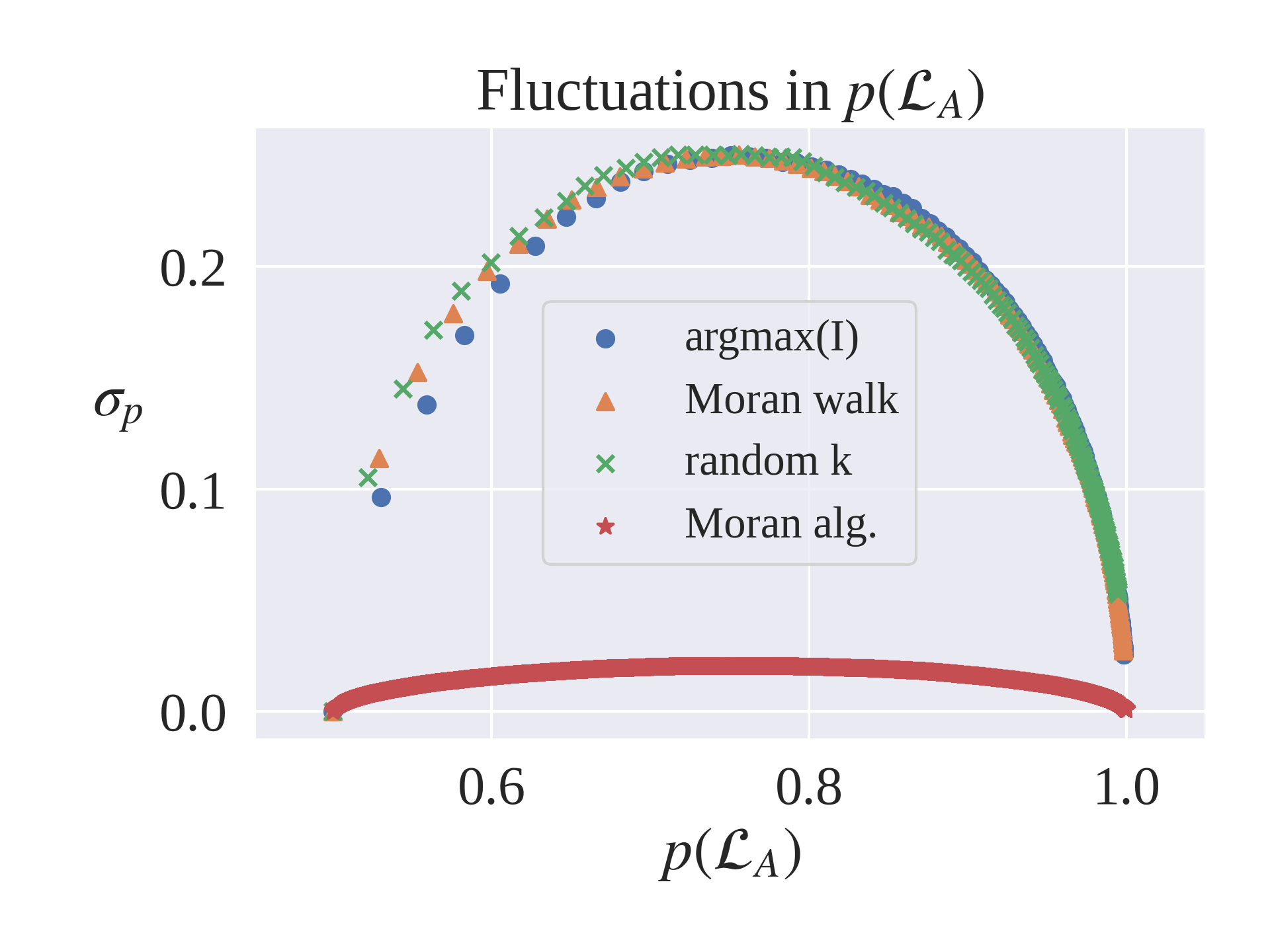}
%\vspace{-0.4in}
    \caption{{\bf \boldmath Confidence fluctuations: } Computed standard deviations $\sigma_p$ for each value $p=p(\Fspace_A)$ of the priors for each algorithm, taken over $\mathcal{N}=12,000$ trials. We use $k/N$ as the prior for the Moran algorithm and compute the standard deviation of $k/N$ across the $12,000$ trials.} \label{fig:fluctuations}
\end{figure}

FIGURE~\ref{fig:slow_and_steady} also shows that although taking much longer to converge, the Moran algorithm has substantially reduced variation in this measure of confidence, compared to fluctuations in the priors in the Bayesian approach. We quantify these fluctuations in FIGURE~\ref{fig:fluctuations} by plotting the standard deviation of the priors in each algorithm as a function of the prior itself (this is possible since the ensemble-averaged prior is found to be monotonic in time). We use $k/N$ as the ``prior" over $\Fspace_A$ in the Moran algorithm. We treat the standard deviation as a function of $p=\bar p_t=\frac{1}{\mathcal{N}}\sum_i p_t^i$, where $p_t^i$ is the prior of the $ith$ trial at step $t$ ($p_t^i=k_t^i/N$ in the case of the Moran algorithm). The standard deviation is computed by,
\be
\sigma(p=\bar p_t) = \sqrt{\frac{1}{\mathcal{N}}\sum_{i=1}^\mathcal{N}(p_t^i-\bar p_t)^2}
\ee

Figures~\ref{fig:slow_and_steady} and \ref{fig:fluctuations} suggest that evolution takes a slower but steadier approach to learning, compared to the faster, but fluctuating, Bayesian approach. 

\subsection{$\log\left(p_{error}\right)$ is the deficit from an optimal reward}

We have compared these learning algorithms in their capacity for extracting information from the landscape, but we can also consider a game-theoretic scenario where each algorithm is assigned a payoff value. We will show that the payoff an algorithm accumulates can be linked to the rate of information extraction that we have discussed above, namely to the reduction of $p_{error}$. A simple reward scheme assigns a payoff $Q_A$ whenever an algorithm decides on type $A$ and payoff $Q_B<Q_A$ when deciding on type $B$, assuming as usual that $\Fspace_A$ is set as the true landscape. The average payoff at time $t$, averaged over an ensemble of $\mathcal{N}$ trials, is quantified by $\bar Q(t) = p_{error}(t) Q_B + (1-p_{error}(t)) Q_A$, where $p_{error}(t)$ as before is the fraction of samples that the prior (or population distribution) is below the $0.5$ mark at a given time $t$; i.e. it is the probability of deciding on $\Fspace_B$, the incorrect landscape. As each algorithm gains certainty that landscape $\Fspace_A$ is correct this average payoff tends towards $Q_A$, the maximal payoff. And so we can quantify the deficit from the optimal payoff for each algorithm at time $t$ with the following,
\begin{align*}
    \log\left(Q_A - \bar Q(t)\right) = \log\left(Q_A-Q_B\right) + \log\left(p_{error}(t)\right). 
\end{align*}
We see that up to a constant that depends on the arbitrary choice of the payoffs $Q_A$ and $Q_B$, the probability of error $p_{error}(t)$ measures the extent to which an algorithm reaches the optimal payoff. And so it is the rate of information extraction that limits the payoff that could be gained from an algorithm. FIGURE~\ref{fig:p_e} thus is equally well interpreted as showing the rates that the algorithms reduce their deficit from an optimal payoff.

\section{Discussion}
We have proposed an analogy between communication channels and the information exchange between an
environment and an evolving population that inhabits it. In the present work we have assumed that
the environment, which specifies a fitness landscape for the evolving population, remains constant
in time and that the particulars of the landscape, namely the superior genotype, are communicated
to the population in the form of birth/death processes. Our interest
has been to understand how this kind of information flow can communicate the
distinction between landscapes, and for our first pass of this problem we consider the absolute
simplest paradigm, just two landscapes $\{\Fspace_A,\Fspace_B\}$. We developed the
notion of a $\Delta$-channel, representing the information transmitted to the population by the
environment in the form of 1-step population changes, $\Delta_k$. We used the mutual information $I(\Delta_k;\Fspace)$ to quantify the average information delivered in this process. We saw that the population changes provide varying
degrees of information depending on the population distribution $k/N$. This inspired a Bayesian
algorithm that made use of the points in the mutual information curve that are most
informative. This method uses the population distribution to probe the environment in an efficient manner. We found that this algorithm performs better than a naive pick-$k$-randomly approach,
or an online approach that uses the natural course of the Moran process (a ``Moran walk") to probe the landscape. Comparatively, evolution, modeled by what we have called the Moran algorithm, is relatively slow in developing confidence in genotype superiority. As we have suggested earlier, this is not especially surprising since natural selection is not optimized for information extraction so much as it is for fitness ascension. Nevertheless, our work positions evolution in the context of other learning algorithms to evaluate its effectiveness in learning an environment. In particular we find what evolution is not: it is not an optimal procedure for extracting information associated with genotype superiority in an environment.

We also found that although the Bayesian algorithm can detect the correct landscape orders of magnitude faster than the Moran algorithm, the confidence with which it decides on the landscape, measured by the values of the priors over the landscapes, fluctuates substantially more than a comparative measure of confidence in the Moran process, namely the population distribution itself. Insofar as the Moran process can be said to properly model evolution, our analysis suggests that evolution represents a slow but careful (low fluctuating) approach to gaining
certainty in genotype superiority. Whereas, algorithms like our Bayesian approach demonstrate that learning the superior genotype can occur much faster than evolution, but at the expense of significantly fluctuating confidence. We speculate that such a slow but low-fluctuating
approach in the case of evolution may be to the advantage of a population attempting to persist in
an environment. The hesitation with which the evolution decides on the superior
genotype saves a population from committing too soon
to a possibly erroneous prediction. 

We also showed that in the context of a simple reward system, it is the probability of error that limits the payoff an algorithm achieves. Though straightforward, this underlines the significance of information extraction as a fundamental mechanism for exploiting the environment. An algorithm that can extract information more rapidly can exploit that information more rapidly as well. 

However, these algorithms have only operated on a static environment. We may find that although evolution is not optimal for information extraction, it may be optimal for achieving persistence in an environment, especially in fluctuating environments. And so we would also like to understand how our learning algorithms fare in time-dependent environments as well. For one thing, our communication channel analogy is more cohesive in a fluctuating setting since the string of environmental states (landscapes) is more analogous to a signal transmitted over the $\Delta$-channel. Only a single environmental state was transmitted in the present work. Considering fluctuating environments will also
allow us to make contact with many important results from the game-theoretic approach to evolutionary biology \cite{Kussell,Villa_Martin} as well as
other approaches to information-acquisition by evolving populations \cite{Bergstrom,McGee}.

We may also like to explore any potential utility in introducing a threshold $\varepsilon >0$ around the $0.5$ mark so that landscape $\Fspace_A$ is selected only if $p_t > 0.5 + \varepsilon$, and landscape $\Fspace_B$ is selected if $p_t < 0.5 - \varepsilon$. And if $0.5-\varepsilon \le p_t \le 0.5 + \varepsilon$, we flip a coin, say. Our Bayesian approach effectively set $\varepsilon=0$. We may find that we can optimize $\varepsilon$ to reduce the probability of error further. 

Lastly, we are
also curious to explore connections to rate-distortion theory in the tracking problem
that a population performs in a fluctuating environment. In particular, we would like to understand the manner in which the entropy rate of the fluctuating environment and any constraints generated by the channel itself place fundamental restrictions
on a population's ability to track and exploit an environment.

\clearpage
\newpage

\newpage
\bibliographystyle{plain}
\bibliography{refs}

\end{document}